\documentclass[10pt,aps,amsmath,showpacs,amssymb,nofootinbib,pre,epsfig]{revtex4}
\usepackage{graphicx} 
\usepackage[usenames,dvipsnames]{color}
\usepackage{float}
\usepackage[normalem]{ulem}
\usepackage{adjustbox}
\newcommand{\blue}[1]{\textcolor{black}{#1}}

\newcommand{\mage}[1]{\textcolor{black}{#1}}
\newcommand{\red}[1]{\textcolor{black}{#1}}
\newcommand{\stk}[1] {\ifmmode\text{\sout{\ensuremath{#1}}}\else\sout{#1}\fi}

\begin{document}

\title{Scale-free behavior of weight distributions of connectomes}
\author{Michelle Cirunay (1), G\'eza \'Odor (1), Istv\'an Papp (1) and Gustavo Deco (2)}
\address{(1) Institute of Technical Physics and Materials Science, Center for Energy Research, P. O. Box 49, H-1525 Budapest, Hungary \\ 
(2) Center for Brain and Cognition, Theoretical and Computational Group,
Universitat Pompeu Fabra / ICREA, Barcelona, Spain \\
}
\maketitle

\section*{Abstract}
To determine the precise link between \red{anatomical structure} and function, brain studies primarily concentrate on the anatomical wiring of the brain and its topological properties. In this work, we investigate the weighted degree and connection length distributions of the KKI-113 and KKI-18 human connectomes, the fruit fly, and of the mouse retina. We found that the node strength (weighted degree) distribution behavior differs depending on the considered scale. On the global scale, the distributions are found to follow a power-law behavior,
with a roughly universal exponent close to 3. However, this behavior breaks at the local scale as the node strength distributions of the KKI-18 follow a stretched exponential, and the fly and mouse retina follow the lognormal distribution, respectively which are indicative of underlying random multiplicative processes and underpins non-locality of learning in a brain close to the critical state. However, for the case of the KKI-113 and the H01 human (1mm$^3$) datasets, the local weighted degree distributions follow an exponentially truncated power-law, which may hint at the fact that the critical learning mechanism may have manifested at the node level too.

\section{Introduction}

In neuroscience, networks-based analysis is heavily needed as the brain is a complex system made up of many interacting neurons~\cite{BarabasiNEUROSCI2023}. Understanding the structure of these connections is crucial as the form is believed to be highly linked with function~\cite{KristanCURBIO2006}. The objective description of the nodes' and edges' contributions to the network as a whole is made possible by the network metrics. Regardless of how widely dispersed or how closely spaced apart they are, nodes with similar qualities can be grouped into a single structurally defined class~\cite{SpornsNETB2016}. 

In literature, one may find many attempts to investigate how the morphological and topological quantities \mage{are }related to neuronal development. For example, earlier works on the structural neural circuits of the cerebral cortex characterized the cortex as a mixing device whose cortico-cortical connections are primarily determined by chance and may be further refined during the learning process~\cite{BraitenbergCORTEX2013}. An even older model of neural networks, which Beurle thought might be shaped by learning and plasticity, was based on random connectivity, an unstructured substrate, in parallel with these neuroanatomical concepts~\cite{BeurlePHILO1956}. Recently models were proposed, in which network optimization is taken into account\mage{. F}or example, Lynn et. al.~\cite{Lynn2024} proposed a model in which following a random edge pruning new link is added either by a preferential attachment or randomly. This provides heavy-tailed connection strengths in agreement with the connectomes they considered. Others have even looked at combined topological and spatial properties of neural networks by making an analogous physical construct of connectomes, which they called \textit{contactomes}~\cite{KovacsNEUCOG2024}, and found that optimization with certain boundary conditions leads to degree and distance distributions close to the neural connectomes in the fruit fly, mouse, and human and unveil a simple set of shared organizing principles across these organisms. While neuronal distance distributions are known to follow exponential rule~\cite{RavaszNEU2013}, degree distributions exhibit more fat-tailed like distribution tails, typically stretched exponential~\cite{GastnerSCIREP2016}. Scale-free behavior was found at the global level of weights in case of large human white matter bundles~\cite{OdorPRE2016} and in case of neural links of the hemibrain~\cite{OdorPRR2022}.

\mage{Brain criticality hypothesis states that the brain activity persists between periods of rapid extinction and amplification~\cite{BeggsFRONTIERSPHYSIO2012}. According to theoretical and experimental evidences, the brain functions close to this region~\cite{BeggsJNEURO2003, HilgetagPHILOTRANS2000, ShewNEUSCI2013, HaimoviciARXIV2013,Hahn-2017} which optimizes its computational capabilities~\cite{BeggsFRONTIERSPHYSIO2012}.
In critical dynamics, a common marker of criticality in a system is the presence of power-law distributed quantities. Previously, it has been hypothesized that functional brain networks follow scale-free behaviors characterized by the presence of power-laws~\cite{EguiluzPRL2005,HeuvelNEUROIM2008}. Power-laws signal a certain amount of self-organization, either through replication (as in biological/metabolic networks) or growth and preferential attachment (as in sociological/technological networks)~\cite{BarabasiSCIENCE1999,PiekniewskiIEEE2009}. Moreover, it has been proposed that the so-called \textit{scale-free} property enables effective communication using a limited number of core nodes that serve as information flow hubs such as in the case of transportation networks~\cite{ZuccaBIORX2019}. Although there are many possible mechanisms that produce power-laws\blue{~\cite{TouboulPRE2017}}, critical behavior optimizes information processing. Power-laws can therefore be used as a tool to look into the criticality of neural systems data~\cite{TinkerSYSNEURO2014}.}

\mage{One of the main neural processes that occur during a living organism's lifespan is learning. Learning takes place because of the alterations in the strength and number of connections that happen between existing neurons. Additionally, frequently used pathways are reinforced and decay with inactivity more formally called Long-Term Potentiation (LTP) and Long-Term Depression (LTD), respectively~\cite{OwensLIFESCIED2017, ParkKORJPHYS2014}. In this work, we investigate how critical learning mechanisms (function) affect the weight and strength distributions of node connections (form) of human and non-human connectomes on both the global and the local scale. In the absence of availability of baby connectome datasets, we utilize a fruit fly larva and adult fruit fly data to study the developmental changes that has occurred.}

\section{Methodology}
The connectome is defined as the structural network of neural connections in the brain~\cite{SpornsPLOSBIO2005}.  At the size of a single neuron, existing imaging methods are unable to fully resolve the roughly $\approx 10^{11}$ neurons that make up the human brain. In this work, we employed coarse-grained networks, acquired using diffusion tensor imaging, including $\approx 10^6$ nodes which is found to agree well with ground-truth data from histology tract tracing~\cite{LandmanNEUROIM2011, DelettreNETNEURO2019}. Such large, whole-brain network data are obtained from the Open Connectome Project repository and have been previously analyzed~\cite{GastnerSCIREP2016}. \mage{The enormous number of nodes results from the usage of various parcellations that are closer to voxel resolution. For instance, the brain masks of a conventional aligned MRI with 1mm resolution include about 1.8 million voxels.} Here, \mage{various connectome datasets are considered: human} KKI-113, KKI-18, \mage{H01 (1mm$^3$)}, fruit fly \mage{(various versions with properties detailed below)}, and the mouse retina. \mage{All of which are considered large enough to avoid finite-size limits.} Table~\ref{tab:netprops} shows the network properties of these datasets.

\begin{table}
\caption{Properties of the Giant Connected Components (GCC) of the networks considered}
\begin{center}
\begin{tabular}{ |c|c|c| } 
\hline \label{tab:netprops}
Dataset & No. of Nodes & No. of Edges \\
\hline
KKI-113& 799,133 & 48,096,501 \\ 
KKI-18& 797,759 & 46,524,003 \\
H01 Human (1mm$^3$)    & 13,579 & 76,004 \\
Mouse retina & 1,076 & 577,350 \\ 
Fly (Hemibrain) & 21,662 & 3,413,160 \\ 
Fly (Hemibrain reciprocated) & 16,804 & 3,251,362 \\ 
Fly (Full brain) & 124,778 & 3,794,527 \\
Fly (Full brain filtered) & 18,103 & 157,904 \\
Fly (Larva) & 2,952 & 110,677 \\
\hline
\end{tabular}
\end{center} 
\end{table}

The KKI-113 network contains $799,113$ nodes and $48,096,500$ weighted and directed edges. On the other hand, the KKI-18 contains  $836,733$ nodes and $46,524,003$ weighted and directed connections~\cite{OCP, gastner_topology_2016}. Additionally, the human H01 (1mm$^3$) dataset contains $13, 579$ nodes and $76,004$ edges~\cite{KovacsNEUCOG2024}. To serve as a comparison, we also consider a fly's full brain with $124,778$ nodes and $3,794,527$ edges and a filtered version, where cells labeled as GLUT, ACH and GABA were removed following~\cite{LinNATURE2024}, having $18, 718$ nodes and $157, 904$ edges~\cite{down-fullbr}, a fly hemibrain~\cite{down-hemibrain1.0.1} (with $21,662$ nodes and $3,413,160$ edges) connectomes and its bidirectional version (with $16, 804$ nodes and $3, 251, 362$ reciprocal edges). Additionally, to investigate the changes in a fly's brain, we included a fly larva dataset~\cite{Larva-fly} with $2,952$ nodes and $110,677$ edges. Finally, we also include the mouse retina data set, with only $1,076$ nodes and $577,350$ weighted and directed edges~\cite{neurodata}. \mage{The neural mouse retina has the interesting property of exhibiting homeostatic plasticity after the development stages characterized by the remarkable ability to maintain a stable architectural and functional organization~\cite{StrettoiINTJMOLSCIE2022}. Here, it would be interesting to see how its edge weights and node strength distributions compare with other organisms that exhibit more flexible plasticity over time as a to response external stimuli.}

\mage{As mentioned in the previous section, we take particular interest in the global and local weight distributions of in the datasets considered.} The global node strength $w_i$ refers to the number of edges that surround a particular node $i$. In weighted networks, the node strength is the generalization of the node degree, or how strongly a node is connected to the rest of the nodes in the networks~\cite{BarratPNAS2004}. The weighted node out-degree \blue{(node strength)} $s_{i}^{out}$ is the sum of the edge weights of outgoing edges
\blue{emanating from} node $i$~\cite{BarratPNAS2004}:
\begin{equation}
\label{eqn:out-degree} s_{i}^{out} = \sum_{j} a_{ij}w_{ij} \ .
\end{equation}
\blue{Here, the weights $w_{ij}$ correspond to the number of links between two nodes in the network 
and $a_{ij}= 0,1$ are adjacency matrix elements, describing the connections between nodes $i$ and $j$}.
\noindent Similarly, the weighted in-degree/strength $s_{i}^{in}$ is the sum of the incoming edge weights for links incident to that node:
\begin{equation}
\label{eqn:in-degree} s_{i}^{in} = \sum_{j} a_{ji}w_{ji}   \ .
\end{equation}

 \mage{Additionally, we also present results on the distribution of the voxel Euclidean distances of the largest human connectome, the KKI-113 in addition to the previously reported topological properties for other human connectomes~\cite{GastnerSCIREP2016}.}

\section{Results}

\subsection{Neuronal Distances}

\begin{figure}[H]
    \centering
    \includegraphics[width=0.8\textwidth]{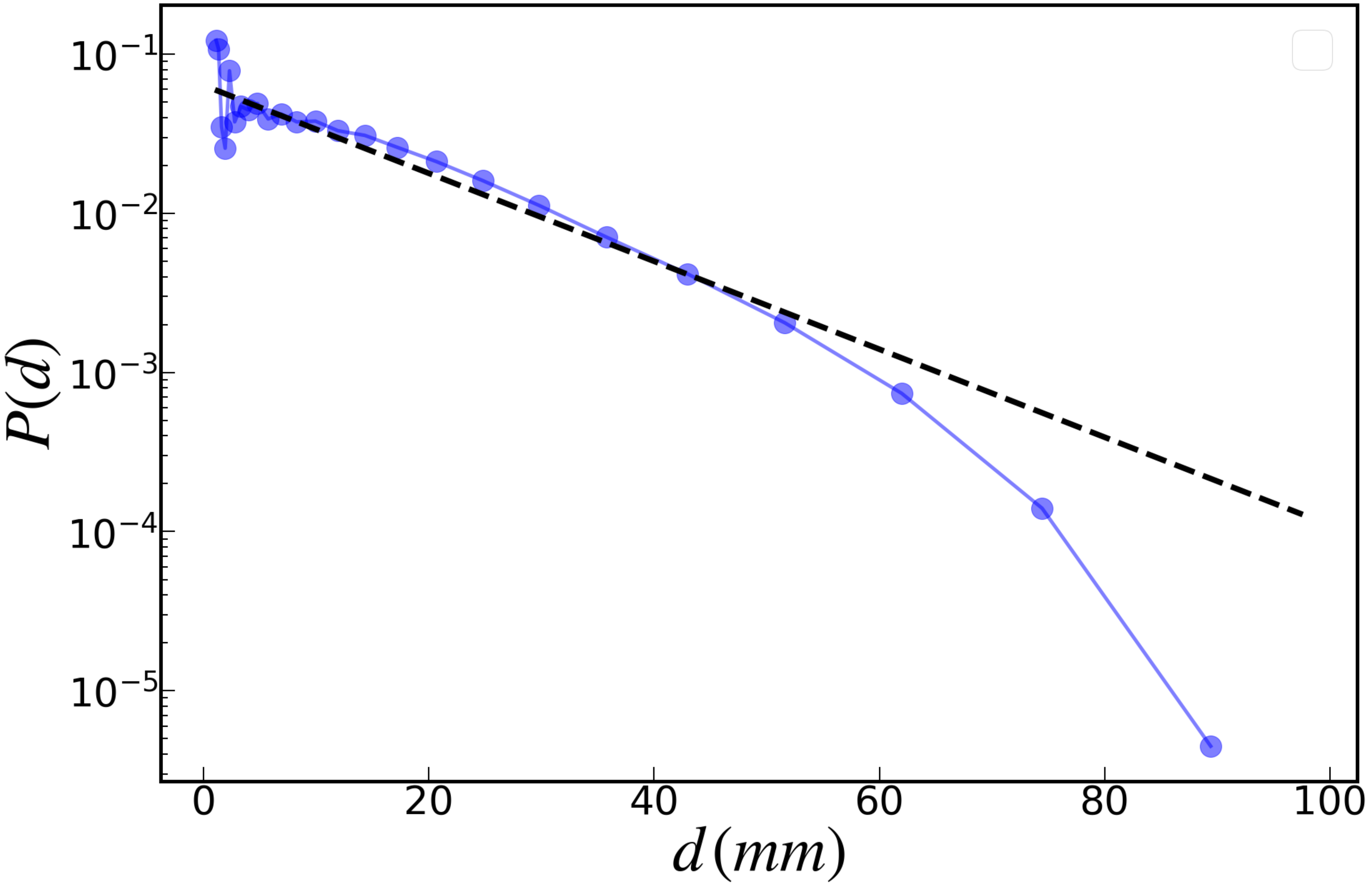}
    \caption{\blue{White matter fiber tract lengths of the KKI-113 are in accordance with the exponential law with a distance scale $d_0 = 15.72$ mm
     and cross over to an even faster decay for $d > 40$ mm.}}
    \label{fig:neuron-dist}
\end{figure}

To describe the spatial arrangement of KKI-113, we regard nodes, whose location coordinates $(x,y,z)$ correspond to its center.
Here, we computed the Euclidean distance between them and obtained the distribution displayed in Figure~\ref{fig:neuron-dist}. In literature, neuronal distances are known to follow the exponential rule $p(d) \propto e^{-d/d_0}$ \mage{for the case of the inter-areal cortical network connections in macaques, mice, and rats ~\cite{RavaszNEU2013}. Furthermore, there is an evidence that the probability of local connections also decays exponentially~\cite{MarkovCERCOR2014, KurthBIORXIV2024}.} Previous works have hypothesized that the establishment and maintenance of synapses in neural connectomes is connected to wiring cost, which aligns with the concept of exponential decay~\cite{AhnPHYSA2006, BullmoreNATREV2012, ChenPNAS2006}. By visual inspection, the distribution of the neuronal distances for KKI-113 seems to agree with such a concept with a characteristic value of $d_0 = 15.72$~mm corresponding to the mean of the data. As one may observe the tail is found to decay faster. This may be due to the limitations of the Diffusion Tensor Imaging which was employed for data acquisition wherein there is a possibility of underestimating long-distance connections~\cite{LiNEUROIM2012, WedeenNEUROIM2008} and overestimating local connections~\cite{JbabdiBRAINCON2011} as in the case of the typical global tractography approach in which streamlines weighted with their corresponding fiber lengths are traced to connect pairs of given voxels~\cite{FingerPLOS2016}. Although \mage{a} more recent work on the use of diffusion MRI tractography and histological tract-tracing applied on ferret brain has shown good agreement between anatomical experiments and the estimates done for the case of mouse and monkey~\cite{DelettreNETNEURO2019} increasing confidence in the technique, we believe that the observed faster decay for the case of KKI-113 was due to this imaging limitation and the fact that the human brain is far more complex and contains more white matter than any other nonhuman primates~\cite{SchoenemannNATNEURO2005}, which \mage{may have introduced} difficulty in delineating neuronal connections, especially at large distances. Moreover, it has been suggested that in larger brains, long-range connections may require greater axon diameters in order to sustain fast neural transmission~\cite{KarbowskiBBIO2007, RingoCERCOR1994} and that certain types of high-cost connectivity may be less common in larger brains~\cite{PhillipsPRSB2015}.

\subsection{Global weights and local strength distributions}

\begin{figure}[H]
    \centering
    \includegraphics[width=\textwidth]{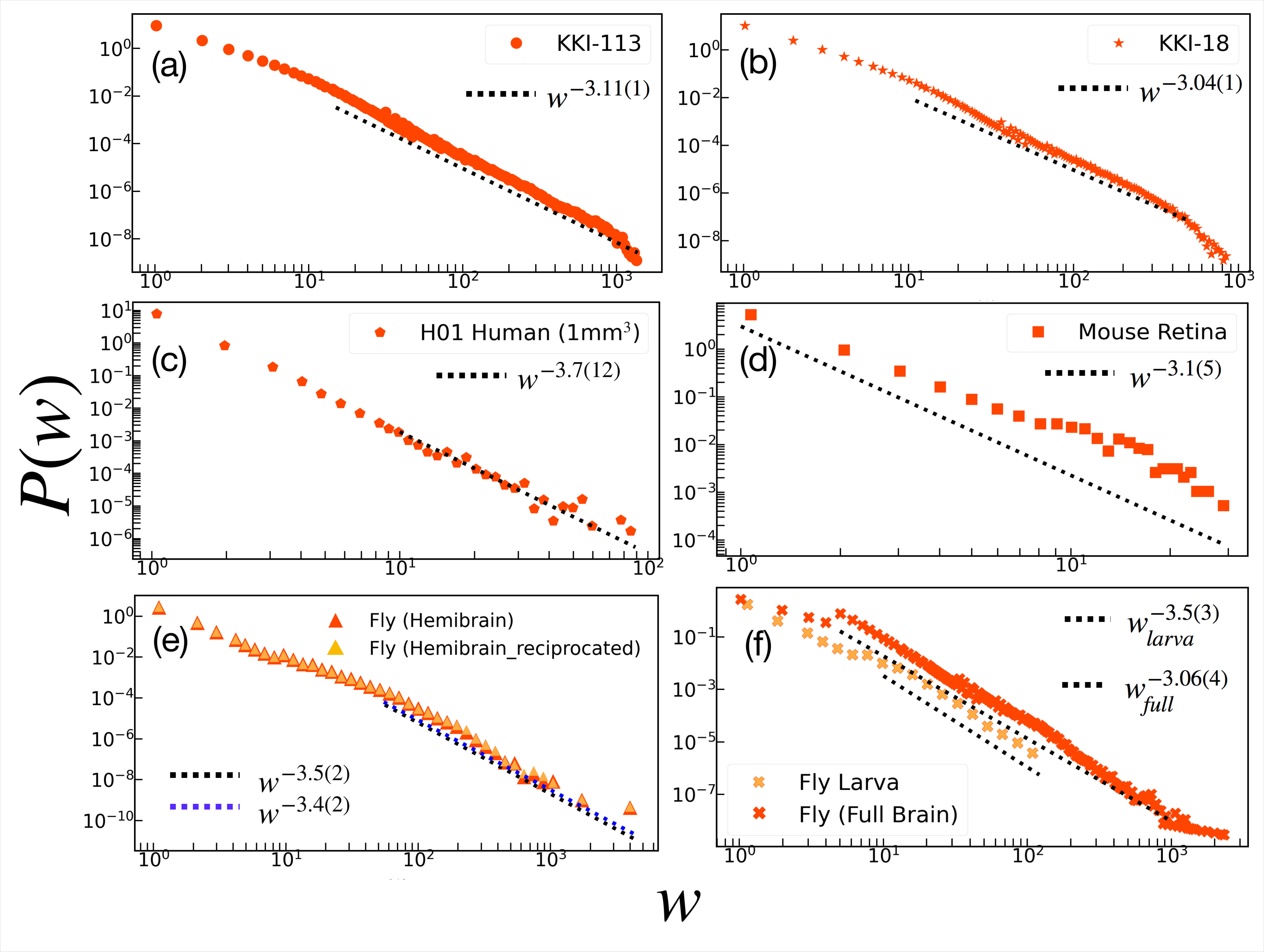}
    \caption{The weight distributions of (a) KKI-113 (b) KKI-18 (c) H01 Human (1mm$^3$) (d) Mouse Retina (e)Fly (Hemibrain) (f) Fly (Full Brain, and its filtered version) and that of a larva's all follow power-law behaviors in the global scale.}
    \label{fig:global}
\end{figure}

\begin{table}
\caption{Global weight distributions fitting parameters}
\begin{center} \label{tab:globalfit}
\begin{tabular}{|l|r|l|}
\hline
Dataset   & Power-Law exponent, $\alpha$ & KS Distance, $D$  \\
\hline
KKI-113     & 3.11(1) & 0.040 \\
\hline
KKI-18     & 3.04(1) & 0.052 \\
\hline
H01 Human (1mm$^3$)    & 3.7(12) & 0.057 \\
\hline
Mouse Retina        & 3.1(5) & 0.087 \\
\hline
Fly (Hemibrain) [FHB] & 3.5(2)    & 0.013 \\
\hline
Fly (Hemibrain reciprocated) [FHBR] & 3.4(2)    & 0.019 \\
\hline
Fly (Full Brain) [FFB]  & 3.06(4)  & 0.030 \\
\hline
Fly (Full Brain filtered) [FFBF] & 3.0(6) & 0.17 \\
\hline
Fly (Larva) [FL] & 3.5(3)  & 0.070  \\
\hline
\end{tabular}
\end{center} 
\end{table}


\mage{Recent works have shown that the computational complexity of a single biological neuron is equivalent to that of 5-8 layers of deep neural networks (DNN) which they attributed to the tree-like morphology of their dendritic branches~\cite{BeniaguevNEURON2021} as well as to their electrical properties~\cite{MillerNATURE1990, BrunelCUROPNEURO2014}. The brain as a complex adaptive system is made up of these individual computational units, whose interactions (i.e. spike-timing dependent synaptic plasticity (STDP)~\cite{EffenbergerPLOS2015}, Hebbian Rules~\cite{DeArcangelisPNAS2010}, short-term plasticity~\cite{LevinaNATPHY2007}) can be considered endogenous factors of neural dynamics that drive the brain to criticality~\cite{TianNETNEURO2022}. One common indicator of criticality in a system is the presence of power-law distributed quantities \blue{and the scaling laws among them}. Although there are many other possible mechanisms that produce power-laws~\cite{TouboulPRE2017}, we note that criticality optimizes information processing~\cite{BeggsFRONTIERSPHYSIO2012} and thus, there is reason to believe that in the context of learning, the presence of power-law is driven by such mechanism.} Power-law distribution analysis can therefore be used to look into the criticality of nervous system data~\cite{TinkerSYSNEURO2014}. \mage{In the following, we present how critical learning mechanism influence the global and local weight distributions of the datasets being considered.}

Figure~\ref{fig:global} shows the global weight distributions of the KKI-113, KKI-18, human H01 (1mm$^3$), mouse retina, and the hemi- and full brain of fruit flies, and that of a fruit fly larva. We employed the statistical framework of Clauset et.al.~\cite{ClausetSIAM2009} to identify and measure power-law behavior. \mage{A pure power law allows for arbitrarily large or small values and because we are dealing with empirical data, we can only fit a power law for a limited range of values. Here, we determine the power-law by assuming an $x_{min}$ (where $x_{min} > 0$) and via the behavior of the tail of the distribution.} We checked the goodness-of-fit by computing the Kolmogorov-Smirnov (KS) distance $D$ and the standard error coefficient $\sigma$ \mage{(see Supplementary Material, Section B for the details of Clauset et.al's KS statistics implementation)}. The exponent which minimizes \mage{ the value of the KS distance} is displayed on the Figure. \mage{Here, we find that on the global scale, the
distributions are found to follow a power-law behavior, with a roughly universal exponent close to 3. The scale-free behavior allow for effective communication due to the presence of core nodes that serve as information hubs~\cite{BarabasiSCIENCE1999, ZuccaBIORX2019}.}Note that the global weight distribution of the KKI-18 (Figure~\ref{fig:global}(b)) has been previously investigated~\cite{OdorPRE2016}, with a power-law fit of $\alpha = 3.05(5)$. This was obtained for data with a size of $N = 41,523,931$. This time, however, we obtained a more complete graph with $N = 46,524,003$, such that the power-law exponent varied slightly with an exponent of $\alpha = 3.04(1)$. 

\mage{In the case of the mouse retina (Figure~\ref{fig:global}(d)), we observe that its global edge weights also follow a power-law behavior with exponent of around $\sim 3$ similar to the other organisms. This is despite of the fact that, unlike the other datasets, the mouse retina is not expected to develop much structural changes beyond the developmental of the mouse (homeostatic plasticity) a necessary adaptive mechanism to ensure the stability of the  firing rate in neurons to compensate for prolonged perturbations of neuronal activity e.g. visual and auditory cortices~\cite{TeichertSCIREP2017, TurrigianoCELL2008}}.

\mage{Figure~\ref{fig:global}(e)-(f) show the edge weights distributions of the fruit fly data sets where we found that $\alpha_{full} < \alpha_{larva}$. The larger exponent value for the larva indicating a higher probability of finding nodes that have fewer links or weaker connections to its neighbors can be attributed to the fact} that at this stage, the fruit fly is still in its formative stages of development. Moreover, this state of the larva network where there is a lack of large values of edge weights, may be thought of as its "pre-learning" phase, where neuroplasticity has not occurred yet. Finally, the opposite can be observed for the full brain of an adult fruit fly which has the lowest value of power-law exponent. This means a lower probability of low-value weights and more likelihood of finding high-edge weights as indicated by its fat tail. The presence of high-edge weights may be an indication of the structural and functional reinforcements that have occurred throughout the phases of a fly's development. 

\mage{We also considered a filtered version of the adult fly's full brain where the glia cells are removed. Glia cells are non-neuronal cells that create myelin in the peripheral nervous system, helping maintain homeostasis and providing support and protection for neurons~\cite{JessenNATURE1980}. The volumetric ratios
of glial cells can vary for every species and every region in their brains~\cite{VerkhratskyNEUROGLIA2019}. Inspite of the inconclusiveness on the ratio of neuron-to-glia cells, we still find value in investigating the case of removing glial cells in the fruit-fly as previous works have shown that these cells in the fruit-fly share significant molecular and functional (engulment activity) attributes with mammalian astrocytes~\cite{LoganNEUGLIABIO2007, FreemanBIOPERS2015}. Here, we observe that the exponent of the tails do not vary much with that of the adult fly's full brain and the only difference is that there is now lower probability of finding large edge weights .}

\mage{If we follow this reasoning, one may ask why the global edge weight distribution of the fly hemibrain  which is taken from an adult fruit fly (both the reciprocated and unidirectional datasets),  is steeper than that of a fruit fly larva. For one, by itself, the brain of a fly (\textit{Drosophilia}) is sparse as it is compared to \textit{C.elegans}, a larva zebra fish, and a mouse~\cite{LinNATURE2024}. Additionally, this can be due to structural reasons i.e.  the full brain connectome includes peripheral, visual neurons, while the hemibrain only consists of the central nervous system only. Finally, the data acquisition of considering only a region may have introduced more complication as this time, some nodes and their corresponding connections to other regions of the brain may have been cut as well leading to fewer edges (preponderance of lower edge weight values) in the considered giant connected cluster (GCC)}. 

\begin{figure}[H]
    \centering
    \includegraphics[width=\textwidth]{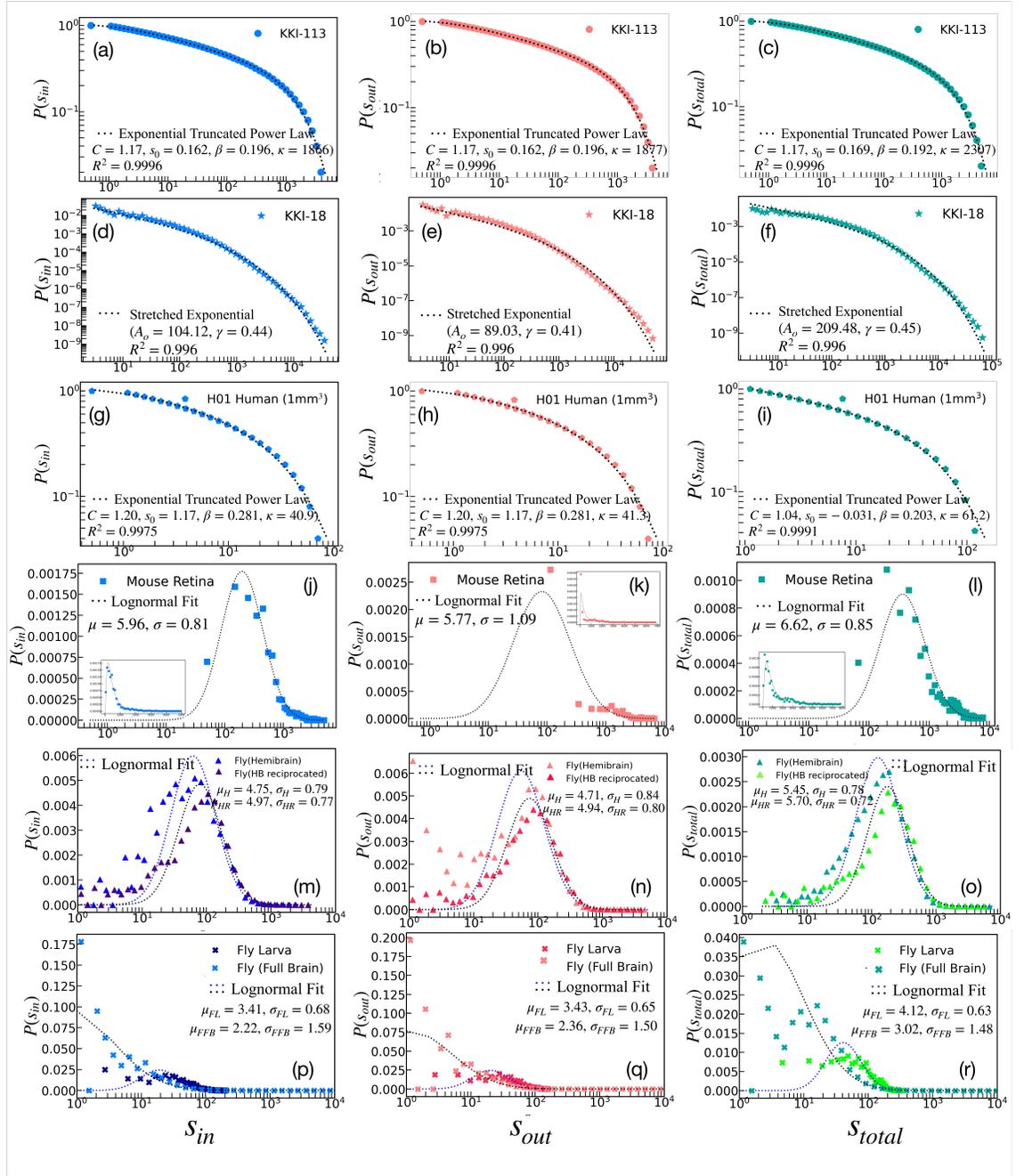}
    \caption{Local weighted degree (node strength) distributions do not follow a power-law behavior anymore. The weighted local in/out-degree of the KKI-113 are shown to behave as an exponentially truncated power-law (shown in (a)-(c)). Meanwhile, KKI-18 follows a stretched exponential ((d)-(f)). The human H01 (1mm$^3$) distributions ((g)-(i)), on the other hand, follow an exponentially truncated power-law. Finally, the weighted degree distributions of the mouse retina ((j)-(l)), and the hemi- ((m)-(o)), full brain ((p)-(r)) of a fruit fly and that of a larva having considerably smaller spatial sizes, all follow lognormal distributions.}
    \label{fig:local}
\end{figure}

\begin{table}[]
\centering
\caption{Local strength distributions fitting parameters}
\begin{adjustbox}{width=\columnwidth,center}
\begin{tabular}{|cc|ccc|}
\hline
\multicolumn{1}{|c|}{Dataset} &
  \begin{tabular}[c]{@{}c@{}}Fitting \\ Function\end{tabular} &
  \multicolumn{3}{c|}{Parameters} \\ \hline
\multicolumn{2}{|c|}{} &
  \multicolumn{1}{c|}{$s_{in}$} &
  \multicolumn{1}{c|}{$s_{out}$} &
  \begin{tabular}[c]{@{}c@{}}$s_{total} = s_{in} + s_{out}$\end{tabular} \\ \hline
\multicolumn{1}{|c|}{KKI-113} &
  Exp. Truncated Power-Law &
  \multicolumn{1}{c|}{\begin{tabular}[c]{@{}c@{}}$C = 1.17$\\$s_o = 0.189$\\ $\beta = 0.150 $\\ $\kappa= 1371$\end{tabular}} &
  \multicolumn{1}{c|}{\begin{tabular}[c]{@{}c@{}}$C = 1.17$\\ $s_o = 0.189$\\ $\beta = 0.150 $\\ $\kappa= 1378$\end{tabular}} &
  \begin{tabular}[c]{@{}c@{}}$C = 1.17$\\ $s_o = 0.189 $ \\ $\beta = 0.147 $\\ $\kappa= 1682$\end{tabular} \\ \hline
\multicolumn{1}{|c|}{KKI-18} &
  Stretched Exp. &
  \multicolumn{1}{c|}{\begin{tabular}[c]{@{}c@{}}$A_o= 104.12$ \\ $\gamma = 0.44 $\end{tabular}} &
  \multicolumn{1}{c|}{\begin{tabular}[c]{@{}c@{}}$A_o= 89.03 $ \\ $\gamma = 0.41$\end{tabular}} &
  $A_o= 209.48$, $c= 0.45$ \\ \hline
\multicolumn{1}{|c|}{H01 Human (1mm$^3$)} &
  Exp. Truncated Power-Law &
  \multicolumn{1}{c|}{\begin{tabular}[c]{@{}c@{}}$C = 1.20$\\$s_o = 1.17$\\ $\beta = 0.281 $\\ $\kappa= 40.9$\end{tabular}} &
  \multicolumn{1}{c|}{\begin{tabular}[c]{@{}c@{}}$C = 1.20$\\$s_o = 1.17$\\ $\beta = 0.281 $\\ $\kappa= 41.3$\end{tabular}} &
  \begin{tabular}[c]{@{}c@{}}$C = 1.20$\\$s_o = -0.031 $ \\ $\beta = 0.203 $\\ $\kappa= 61.2$\end{tabular} \\ \hline
\multicolumn{1}{|c|}{Mouse Retina} &
  Lognormal &
  \multicolumn{1}{c|}{$\mu= 5.96$, $\sigma = 0.81 $} &
  \multicolumn{1}{c|}{$\mu= 5.77$, $\sigma = 1.09 $} &
  $\mu= 6.62$, $\sigma = 0.85$ \\ \hline
\multicolumn{1}{|c|}{Fly (Hemibrain) {[}FHB{]}} &
  Lognormal &
  \multicolumn{1}{c|}{$\mu= 4.75$, $\sigma = 0.79 $} &
  \multicolumn{1}{c|}{$\mu= 4.71$, $\sigma = 0.84$} &
  $\mu= 5.45$, $\sigma = 0.78$ \\ \hline
\multicolumn{1}{|c|}{Fly (Hemibrain reciprocated) {[}FHBR{]}} &
  Lognormal &
  \multicolumn{1}{c|}{$\mu= 4.97$, $\sigma = 0.77 $} &
  \multicolumn{1}{c|}{$\mu= 4.94$, $\sigma = 0.80$} &
  $\mu= 5.70$, $\sigma = 0.72$ \\ \hline
\multicolumn{1}{|c|}{Fly (Full Brain) {[}FFB{]}} &
  Lognormal &
  \multicolumn{1}{c|}{$\mu= 2.22$, $\sigma = 1.59 $} &
  \multicolumn{1}{c|}{$\mu= 2.36$, $\sigma = 1.50 $} &
  $\mu= 3.02$, $\sigma = 1.48 $ \\ \hline
\multicolumn{1}{|c|}{Fly (Full Brain filtered) {[}FFBF{]}} &
  Lognormal &
  \multicolumn{1}{c|}{$\mu= 1.44$, $\sigma = 1.30 $} &
  \multicolumn{1}{c|}{$\mu= 1.39$, $\sigma = 1.35 $} &
  $\mu= 1.95$, $\sigma = 1.35 $ \\ \hline
\multicolumn{1}{|c|}{Fly (Larva) {[}FL{]}} &
  Lognormal &
  \multicolumn{1}{c|}{$\mu= 3.41$, $\sigma = 0.68 $} &
  \multicolumn{1}{c|}{$\mu= 3.43$, $\sigma = 0.65 $} &
  $\mu= 4.12$, $\sigma = 0.63$ \\ \hline
\end{tabular}
\end{adjustbox}
\end{table}

 Figure~\ref{fig:local} shows the local in-, out-, and total (in+out) node strength distributions for all datasets considered. For the case of KKI-113 (shown in Figure~\ref{fig:local}(a)-(c)) and the human H01 (1mm$^3$) datasets, the strength distributions are found to follow an exponentially truncated power-law, also consistent with other large human connectomes~\cite{GastnerSCIREP2016}. \mage{The presence of a power-law, albeit truncated, may hint that the critical mechanism \blue{is} manifested at the node-level too~\cite{AnsellCOMPHYS2024}.} The truncation may be due to some physical constraints. For example, Mossa et. al. attributed the truncation on the degree distributions of the World Wide Web (WWW) and the University of Notre Dame domain, to the limitation in information-processing capabilities of the nodes. As there is a cost associated with information-processing, there was a need to filter incoming information based on \textit{interest}. By doing so, the new nodes in the network only process a subset of the information from existing nodes~\cite{MossaPRL2002}. As the brain itself is an information-processing unit, we believe that this may also explain the truncation in the local weighted degree distributions of the KKI-113 and the H01 (1mm$^3$) datasets, wherein neuronal connections and exchanges only occur with a subset of local neighbors. Additionally, in systems that restrict the maximum number of linkages a node can have, such as the local neighborhood being considered, the scale-free characteristic is not to be expected. The ability of the nodes to link to an arbitrary number of other nodes is necessary for the scale-free property to appear. 

Stretched exponential and lognormal distributions may arise from \textit{multiplicative processes}. In fact, in some cases, Laherre and Sornette proposed the stretched exponential as an alternative to the power-law~\cite{LaherreEPJB1998}. With multiplicative phenomena, one instance can multiply rapidly, triggering a cascade. When something is of this nature, individual instances are not independent of one another. These behaviors can be expected to arise from a high level of connectivity. Here, we observe that such distributions not only describe the possible types of events or cascade that flow through these connections i.e. electrical signal, infection, etc., but this time, they describe the level of connectivity itself (i.e. weighted degree). As shown in Figure~\ref{fig:local}, local node strengths ($s_{in}$, $s_{out}$, and $s_{tot}$) of the KKI-18 [(d)-(f)], the mouse retina [(j)-(l)] and the fruit fly [(m)-(r)], are found to follow the stretched exponential or lognormal distributions, respectively. The former is consistent with a previous work on other large human connectomes~\cite{GastnerSCIREP2016}. We surmise that this multiplicative creation of connections has resulted from \textit{neuroplasticity} or the brain's ability to reorganize functionally and structurally as a response to external stimuli~\cite{GrafmanJCOMMDIS2000, DavidsonNATURENEURO2012, McEwenENDOC2018}. Such connections are the physical reinforcements that were created during the different stages of development of the human, \mage{fruit-fly, and the} mouse. 

The formation of synapses in the brain is thought to have begun with mostly undirected "exploratory" extension, which is followed by selective consolidation and contact dissolution in an early developmental model. The first process might be termed \textit{random}, whereas the second process transmits \textit{specificity} since it is primarily driven by neural activity or biochemical interactions between participating cells~\cite{JontesNEURON2000}. As the brain connections were first thought to be random and later shaped by learning and plasticity~\cite{BraitenbergCORTEX2013, BeurlePHILO1956}, this may explain the observed lognormal behaviors that hint at the underlying random multiplicative processes that have occurred in the system.

It may appear difficult for spatially embedded networks to exhibit scale-free behavior because of the inherent limitations imposed by basic spatial and metabolic constraints~\cite{SpornsNETB2016} on the density and number of connections that may be maintained at any given node. An exponential and exponentially truncated power-law was seen in different spatial networks, including transportation networks~\cite{AmaralPNAS2000, GuimeraPNAS2005}, which we have also observed as in the case of the human connectomes, KKI-113, KKI-18, and H01 (1mm$^3$) as shown in Figure~\ref{fig:local}(a)-(i). \mage{For even more constrained networks such as the case of the mouse \mage{retina} and fruit fly (hemi, larva, adult),  Figure~\ref{fig:local}(j)-(r) show that their node strength ($s_{in}$, $s_{out}$, $s_{total}$)  distributions follow the lognormal distribution, characterized by a unimodal heavy-tailed behavior described by the parameters $\mu$ and $\sigma$ (the mean and standard deviation of the natural logarithm of the strengths), representing a characteristic value and variations in node strengths. For these datasets, the proximity in value of the parameters} may hint at the degree of symmetry or reciprocity of the connections at the local scale. This so-called reciprocity of pathways and connectivity has also been observed in the macaque visual cortex~\cite{FellemanCERCOR1991} and mammalian cortex~\cite{SongPLOS2005}. This is also shown in case of the Full-fly, see~\cite{Dorkenwald2023.06.27.546656}. This may be a manifestation of specificity in the form of link consolidation~\cite{JontesNEURON2000}. Expectantly, the values of such parameters for the total (in + out) strength distributions will be slightly larger.

\begin{table}
\caption{Summary of source and  sink nodes}
\begin{adjustbox}{width=\columnwidth,center}
\label{tab:source}
\begin{tabular}{|l|l|l|r|l|l|l|}
\hline
Dataset   & No. source nodes & Max. $s_{out}$ & $\left<s_{out}\right>$  & No. sink nodes  & Max. $s_{in}$ & $\left<s_{in}\right>$ \\
\hline
KKI-113     & 6,684    & 2,100 &  26.55  & 5,686  & 536 & 18.03  \\
KKI-18    & 2,270 & 1,663 & 52.54    & 2,410   & 1,300 & 53.68  \\
H01 Human (mm$^3$) & 2,109 & 42 & 4.63    & 2,145   & 42 & 4.33  \\
Mouse retina  & 1    & 3,977 & 3,977   & 61   & 913 & 255.08  \\
Fly (Hemibrain) & 0    & - & - & 46   & 250 & 62.20 \\
Fly (Hemibrain reciprocated) & 3    & 17 & 9.67 & 1   & 3 & 3 \\
Fly (Full brain) & 8,565   & 640 & 18.85  & 8,809  & 766 & 18.62 \\
Fly (Full brain filtered) & 2,023   & 31 & 1.84  & 2,370  & 39 & 3.18 \\
Fly (Larva) & 58 & 18 & 7.66 & 43 & 38 & 16.74 \\
\hline
\end{tabular}
\end{adjustbox}
\end{table} 

We further witness the concept of specificity in the development of the fruit fly. Comparing the parameters of a larva~\cite{Larva-fly} and that of a full-grown fly~\cite{Dorkenwald2023.06.27.546656}, we observe that the $\mu_{FFB(R)} < \mu_{FL}$ and that $\sigma_{FFB(R)} > \sigma_{FL}$. The same trend is observed even if we consider the filtered version of the adult fly brain. The decrease ($\mu_{FFB(R)} < \mu_{FL}$) in the characteristic node strength in the adult fly brain may be due to the dissolution of connections due to some metabolic processes and other constraints~\cite{JontesNEURON2000}. The larger characteristic node strength (degree/connections) in the larva may be because it is still in the random exploratory stages of forming neural connections. Additionally, these connection strengths cannot vary much from each other (smaller $\sigma$) as the larva only has very few nodes to connect to with such connections not yet being reinforced due to external stimuli and factors~\cite{GrafmanJCOMMDIS2000, DavidsonNATURENEURO2012, McEwenENDOC2018}. The opposite is true for the full brain of the adult fruit fly which contains more nodes and a larger spatial span of connections.

\subsection{Strength distributions of source and sink nodes in the network}

\begin{figure}[H]
    \centering
    \includegraphics[width=0.8\textwidth]{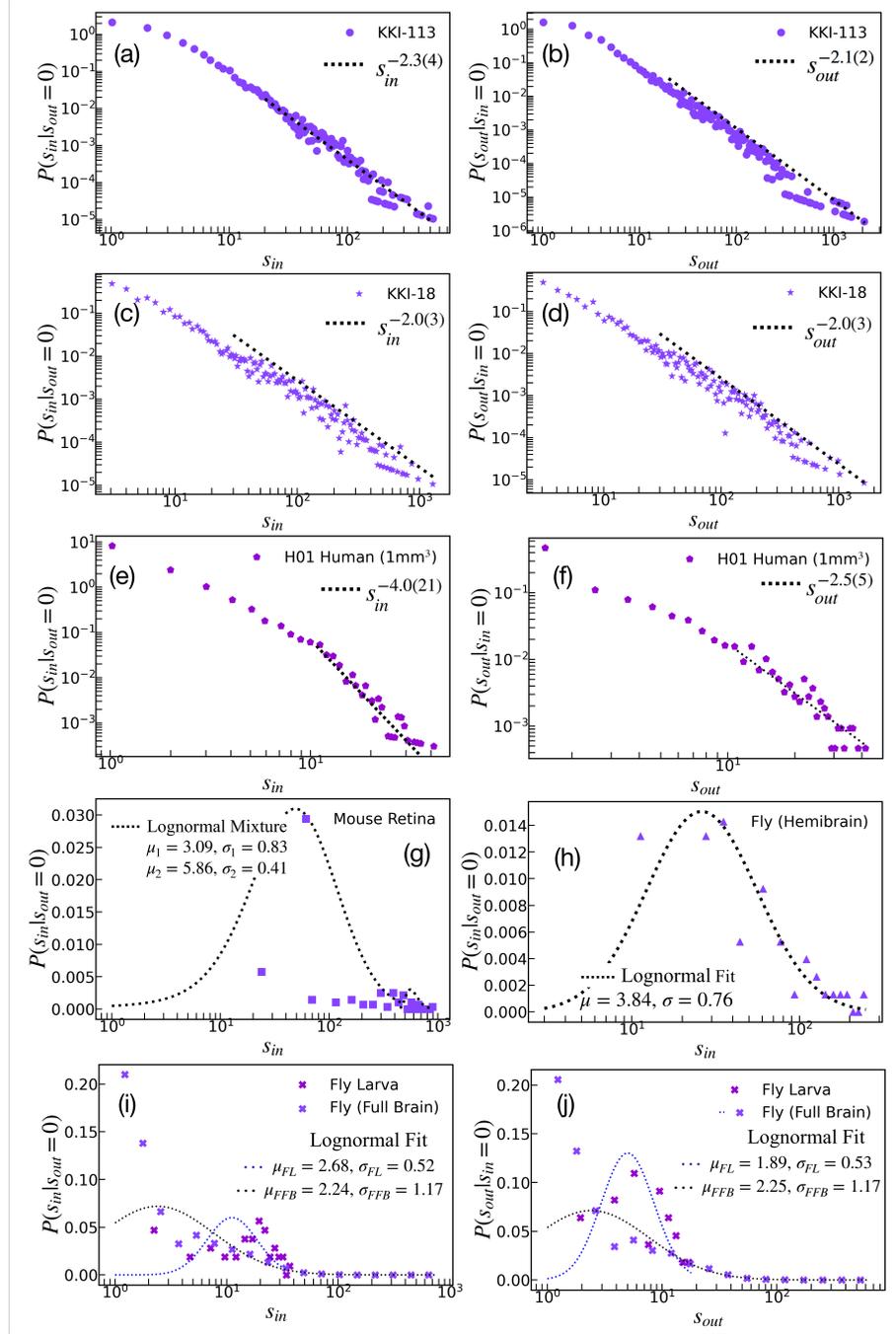}
    \caption{The weighted in and out strengths of source and sink nodes of (a)-(b) KKI-113, (c)-(d) KKI-18, (e)-(f) H01 Human (1mm$^3$), (g) mouse retina (in-strength only), (h) Fly (Hemibrain) (in-strength only) and (i)-(j) Fly (Larva and Full brain).}
    \label{fig:condprob}
\end{figure}

Each network node's contributions to the overall design of a brain network can be measured once it has been defined. The \textit{network participation indices} which \mage{K{\"o}tter} and Stephan studied, showed areas of relatively densely connected nodes that were receiving (referred to as "receivers") and emitting (referred to as "senders") connections~\cite{KotterNEUNET2003}. There are instances when these participatory network metrics they mentioned—like extremely central nodes also have high degree. Information transit may be connected to these network participation metrics~\cite{SpornsNETB2016}. In the following, we explore how the distributions of nodes with purely incoming and outgoing edges contribute to the network. 

Table~\ref{tab:source} summarizes source (nodes with only outgoing edges) and sink (nodes with only incoming edges) nodes with their corresponding average in or out strengths for all the datasets being considered. For KKI-113, (pure) source, and sink nodes make up approximately 0.84 \% and 0.71 \% of the network, respectively. On the other hand, KKI-18 is made up of 0.27 \% source nodes and 0.29 \% sink nodes. The human H01 (mm$^3$) dataset consists of 15.53\% source nodes and 15.80\% sink nodes. For the non-human connectomes, we find cases of either having little to no source or sink nodes, such as the case of the mouse retina (with only one source node) and a fly hemibrain, both uni- and bidirectional edges dataset, contain only a few sink nodes. This may be due to the fact, that the mouse retina and the fly hemibrain, are only sections of an entire organism's neuronal network. Notice, however, that although these non-human connectomes have fewer nodes, they have very high connectivity to the rest of their neighbors as indicated by their relatively high average node strengths. Meanwhile, for a fly larva, being only in its developmental stages, the source and sink nodes comprise 0.02\% and 0.01\% of its entire network. Finally, for the full fruit fly brain, we can observe that its source nodes comprise 6.89 \% of its neuronal network, while its sink nodes make up 7.06\% of it. Its filtered version contains 0.11 \% source nodes and 0.13 \% sink nodes.

 Even though Tables~\ref{tab:netprops} and \ref{tab:source} indicate that these nodes only comprise a small fraction of the networks, some of these nodes serve as hubs, which is evident in the large value of maximum $s_{in}$ and $s_{out}$. Because of this, they can be crucial for neural integration and brain communication, making them important participants in cognitive processes. Furthermore, at times of stress, these network regions are prone to disconnecting and malfunctioning~\cite{HeuvelCOGSCI2013}. For example, mapping the locations of sources and sinks in a mouse \mage{retina} revealed that the prefrontal cortex and other higher-order brain regions are the main neuronal output sources for the rest of the brain, while the basal ganglia are important receivers of incoming projections~\cite{ColettaSCIADV2020}.  In this work, we slightly touch on the topic of source and sink in the neuronal networks of humans (KKI-113, KKI-18), the mouse, and a fruit fly.  This time, we explore the mouse retina and investigate the strength distributions of the source and sink nodes. Additionally, as in the case of humans, sources are found to be high-influence nodes which inhibit the sink nodes from going to epileptic seizures~\cite{GunnarsdottirBRAIN2022}. Figure~\ref{fig:condprob} shows the probability distributions of the node strengths of the source ($P(s_{out} | s_{in} = 0$)) and sink ($P(s_{in} | s_{out} = 0$)) nodes in the network datasets being considered.

Interestingly, for the human connectomes KKI-113, KKI-18, and the human H01 (1mm$^3$) datasets, the strength distributions of these sources and sinks follow power-law tails which have exponents slightly above 2. Note that for the strength distribution of the H01 source nodes, find a sharp knee-point (at $s_{in} \approx 11$) in its distribution followed by a steep tail (with exponent $\alpha \approx 4$). This sharp transition can signal finite-size limitations since we are looking into a 1mm$^3$-sized region.  However, for the case of non-human connectomes (Figure~\ref{fig:condprob}(e)-(h)), we observe lognormal behaviors of source and sink nodes strengths.

\begin{table}[]
\caption{Source-Sink node strength parameters}
\begin{adjustbox}{width=\columnwidth,center}
\begin{tabular}{|l|ll|ll|}
\hline
\multicolumn{1}{|c|}{Datasets} &
  \multicolumn{2}{c|}{\begin{tabular}[c]{@{}c@{}}Sinks\\ ($s_{out} = 0$)\end{tabular}} &
  \multicolumn{2}{c|}{\begin{tabular}[c]{@{}c@{}}Sources\\ ($s_{in} = 0$)\end{tabular}} \\ \hline
 &
  \multicolumn{1}{l|}{Fitting} &
  Parameters &
  \multicolumn{1}{l|}{Fitting} &
  Parameters \\ \hline
KKI-113 &
  \multicolumn{1}{l|}{Power-law} &
  $\alpha = 2.3(4)$, $D = 0.056$ &
  \multicolumn{1}{l|}{Power-law} &
  $\alpha = 2.1(2)$, $D = 0.056$ \\ \hline
KKI-18 &
  \multicolumn{1}{l|}{Power-law} &
  $\alpha = 2.0(3)$, $D = 0.082$ &
  \multicolumn{1}{l|}{Power-law} &
  $\alpha = 2.0(3)$, $D = 0.081$ \\ \hline
H01 Human (1mm$^3$) &
  \multicolumn{1}{l|}{Power-law} &
  $\alpha = 4.0(21)$, $D = 0.039$ &
  \multicolumn{1}{l|}{Power-law} &
  $\alpha = 2.5(5)$, $D = 0.075$ \\ \hline
Mouse Retina &
  \multicolumn{1}{l|}{\begin{tabular}[c]{@{}l@{}}Lognormal\\ Mixture\end{tabular}} &
  \begin{tabular}[c]{@{}l@{}}$\mu_1 = 3.09$\\ $\sigma_1 =  0.83$\\ $\mu_2 = 5.86$\\ $\sigma_2 =  0.41$\end{tabular} &
  \multicolumn{1}{l|}{-} &
  - \\ \hline
Fly (Hemibrain) {[}FHB{]} &
  \multicolumn{1}{l|}{Lognormal} &
  \begin{tabular}[c]{@{}l@{}}$\mu = 3.84$\\ $\sigma = 0.76$\end{tabular} &
  \multicolumn{1}{l|}{-} &
  - \\ \hline
Fly (Hemibrain reciprocated) {[}FHBR{]} &
  \multicolumn{1}{l|}{-} &
  - &
  \multicolumn{1}{l|}{-} &
  - \\ \hline
Fly (Full Brain) {[}FFB{]} &
  \multicolumn{1}{l|}{Lognormal} &
  \begin{tabular}[c]{@{}l@{}}$\mu = 2.24$\\ $\sigma = 1.17 $\end{tabular} &
  \multicolumn{1}{l|}{Lognormal} &
  \begin{tabular}[c]{@{}l@{}}$\mu = 2.25$\\ $\sigma = 1.17$\end{tabular} \\ \hline
Fly (Full Brain filtered) {[}FFBF{]} &
  \multicolumn{1}{l|}{Lognormal} &
  \begin{tabular}[c]{@{}l@{}}$\mu = 0.81$\\ $\sigma = 0.83 $\end{tabular} &
  \multicolumn{1}{l|}{Lognormal} &
  \begin{tabular}[c]{@{}l@{}}$\mu = 0.28$\\ $\sigma = 0.82$\end{tabular} \\ \hline
Fly (Larva) {[}FL{]} &
  \multicolumn{1}{l|}{Lognormal} &
  \begin{tabular}[c]{@{}l@{}}$\mu = 2.68$\\ $\sigma = 0.52$\end{tabular} &
  \multicolumn{1}{l|}{Logrnormal} &
  \begin{tabular}[c]{@{}l@{}}$\mu = 1.89 $\\ $\sigma = 0.53$\end{tabular} \\ \hline
\end{tabular}
\end{adjustbox}
\end{table}

The mouse retina is expectantly smaller with only over a thousand nodes and around half a million edges. By examining the distributions of nodes with either only incoming or outgoing edges, we found that there is only one node in the network that has no incoming edges (identified to be Node $0$ with $s_0^{out} = 3977$). Such voxel may be part of the retina that is closest to the mouse's brain which distributes information to all other neighboring nodes in the network. Additionally, looking at the sink nodes, qualitatively it was observed that after the first peak, there is another rise in the trend before it goes down again. Note that events following lognormal behaviors are independent random variables with mean and standard deviation values that are also independent of each other which means that it does not matter how we regroup our data. Having said this, we closely inspect the data by splitting it into two: $s_{in} \leq 97$ and $s_{in} > 97$. This value was selected because the intermediate bins are empty signaling a decay of the previous mode and the rise of the second modal curve. Here, we saw that each range of data would follow a lognormal trend and as a whole, the entire dataset follows a bimodal lognormal distribution (or lognormal mixture). In literature, such behavior can model the first-order kinetics of chemicals when mixed~\cite{AndersonSCIREP2021} ; subsequent waves of a pandemic~\cite{ValvoAPPSCI2020}; or represent market volatility structures~\cite{BrigoIJTA2002}. In this context, the lognormal mixture observed in the mouse retina dataset implies clustering and differences in the concentration of connections. Here, we observe that there is a high probability of finding sparsely connected regions; and the occurrence of densely connected portions (with more than 100 incoming edges) of sink nodes. According to earlier research~\cite{SongPLOS2005}, synaptic strengths in the rat visual cortex follow a lognormal distribution with a heavy tail, indicating a higher-than-expected abundance of strong synaptic connections.  Furthermore, stronger connections tend to be more heavily clustered~\cite{SpornsNETB2016}, which may account for the datapoint concentration at the lognormal mixture's second peak.

Finally, for the case of the fruit fly datasets (hemibrain, larva, full brain, and filtered full brain), the in and out strengths of the source and sink nodes still obey the lognormal behavior, which may be hinted from the behavior of the local weight distributions shown in Figure~\ref{fig:local}(j)-(l). Similar to the in and out node strength distributions we can still find that the $\mu_{FFB} < \mu_{FL}$ and that $\sigma_{FFB} > \sigma_{FL}$ (shown in Figure~\ref{fig:local}(m)-(o). Notice that this time, the $\mu_{FFBF} < \mu_{FL} $ for both the source and sink nodes. This is because there are more nodes in the filtered version of the adult full fly's brain ($N=18, 103$) still as compared to the larva's ($N = 2,952$). In any case, we believe that the same underlying mechanisms are responsible for these observations which led to them following the same statistical behaviors.

\section{Possible explanation for the power-laws of global weight distributions}

Long-time learning and memory were shown to be induced via the long-term potentiation (LTP) synaptic plasticity, which is related to the longevity \blue{($\tau$)} of neighboring neuron pair firing activity~\cite{10.1093/brain/awl082}. LTP results from coincident activity of pre- and post-synaptic elements, bringing about a facilitation of chemical transmission, that can persist for periods of weeks or months.

\blue{
According to this we assume that the weights $w_{ij} \propto \tau$ and their PDF-s are also proportional to the correlated activity
durations $p(w_{ij}) \propto p(\tau) \propto \tau^{-\alpha}$.
Note, that a weight decrease in case of neurons not firing together should also be taken into account to model forgetting and to avoid 
unbounded growth of $w_{ij}$, but we assume it to be a random process that does not affect the scaling behavior.  
The correlated activity duration distributions, can be related to the auto-correlation functions of
the variables, which exhibits the asymptotic scaling: $C_{AA}(t)\propto t^{-\lambda/Z}$, where $\lambda$ is the auto-correlation 
exponent and $Z$ is the dynamical exponent of the critical process~\cite{odorbook}. 
Here consider a pair of active nodes, which in uncorrelated system exhibits $p_{AA}(t) = p_A(t)^2$, but right at the critical 
point $p_{AA}(t) \propto p_A(t)$~\cite{Schram_2012}. 
A time integral of the two-point (pair) auto-correlations provides the duration PDF-s in the $\tau\to\infty$ asymptotic limit as  
\begin{equation}
    \int_{t_0}^{t_0+\tau} C_{AA}(t)  \mathrm{d} t = p(\tau) \,,
\end{equation}
where $t_o$ is the initial time of the avalanche, giving rise to $p(\tau)\sim \tau^{-\alpha} = \tau^{-\lambda/Z+1}$ and 
an exponent relation $\alpha = \lambda/Z-1$. }

\blue{
For a generic universality class considered to describe brain criticality~\cite{BeggsJNEURO2003}, the Directed Percolation (DP) in the 
high dimensional mean-field limit, the auto-correlation function decays asymptotically with $\lambda/Z = 4$~\cite{henkel2008}, thus 
$p(w_{ij}) \propto p(\tau) \propto \tau^{-3}$ can be expected, in agreement with our measurements for global weight distributions. 
This scaling can be true for the mean-field behavior of several other basic universality classes, like DP-C (Manna) or the 
dynamical percolation~\cite{henkel2008}.
In lower spatial dimensions the autocorrelation exponent decreases, for example in $d=2$ it is $\lambda/Z = 2.5(3)$~\cite{henkel2008}, 
providing an estimate $\alpha = 1.5(3)$ for the two dimensional DP class. 
Model simulations assuming Manna sandpiles are under way for direct confirmations~\cite{COBcikk}. 
We assume, that for white matter fiber tracts (i.e in case of KKI-18) the aggregate of the neuron level learning rules
results in similar distributions.}

However, it is still an open question if the critical brain would belong to the mean-field DP universality class or to other and if non-universal scaling~\cite{MArep, CCrev}, corresponding to a Griffiths phase~\cite{Griffiths} or to an external drive~\cite{PhysRevX.11.021059,Odor_2023}. 
In the case of the Shinomoto-Kuramoto model, possessing periodic external forces, to describe the task phase of the fruit-fly brain 
it was shown that PDF-s of the \blue{inter-event times} decay with non-universal power-laws, characterized by exponents 
\blue{$2 < \nu < 4$}, depending on the strength of the excitation~\cite{Odor_2023} and on the \blue{actual communities}.
Similar behavior was found for the Hurst and beta exponents, which describe auto-correlations. This community dependence has 
been confirmed via fMRI and BOLD signal measurements~\cite{Ochab_2022} in case of humans.

\section{Summary}

In summary, we have investigated the weight distributions of various human and animal connectomes, to search for power-law-tailed PDFs, which can be related to the learning mechanism of the brain in a critical state. These are the largest, publicly available neural graphs.
We found that the global weight distributions can be fitted by power-laws tails, characterized by exponents slightly above 3, the node strength PDF-s decay faster, stretched exponentially, or via lognormal way. The whole brain human fiber tracks, obtained by MRI with deterministic tractography, using the Fiber Assignment by Continuous Tracking algorithm methods~\cite{mori1999three} exhibit the most power-law-like behavior, even the node strengths of the source and sink nodes are like that.

We have compared these results of adult (human and animal) connectomes with that of an untrained, larva fly and found that the latter has a narrow PDF, that starts in the same way as the adult fly. This suggests that a certain degree of learning happens already in premature brain development or the initial axon growth mechanisms could also lead to fat tails in the connection strengths~\cite{Lynn2024}. This should be studied more, using other connectomes obtained in the early phase of neural growth.

We provided a possible model and a scaling relation, which connects the critical auto-correlation function to the LTP plasticity mechanism, that can explain the power-laws we observe. Thus we rely on the mutual relationship between structural and functional connectomes and claim
that non-local learning is dictated by a brain state close to criticality~\cite{Angiolelli2024.07.15.603226}. \blue{Assuming mean-field critical
behavior of the branching processes we derive a global weight exponent $\alpha=3$, which is close to most of the values we obtained by analyzing 
different connectomes.
But in the lack of the true whole-brain function model, we can't give a precise weight exponent estimate. This can also vary across
modules following different dynamics.} Note, that learning, via link strengthening makes the connections more asymmetric, which enhances the violation of the fluctuation-dissipation and makes the brains more non-equilibrium~\cite{Monti2024.04.04.588056}.

Finally, we have also determined the white matter fiber tract distance distributions of the large KKI-113 human connectome. We found that it follows the exponential rule up to 4 centimeter size, but breaks down and decays faster for longer lengths.
Note, that the distances are calculated from the x,y, and z coordinates of nodes as we don't know the real wire lengths. 

\red{Overall, the exploration of the relationship between anatomical structure and function across species reveals scale-free properties at the global level and diverse local-scale behaviors of the node structural degree tied to critical dynamics. Our results suggest that human neural architecture achieves a balance between efficiency and adaptability through a dynamic interplay of global and local properties. While highlighting unique patterns in human connectomes, the study situates these within broader principles shared across species, offering insights into the structural basis for learning, self-organization, and adaptability. This findings help us to elucidate the key question about what makes us humans.}

\section*{Acknowledgements} 

We thank the helpful discussions to I. A. Kov\'acs, decoding of the KKI-113 coordiantes to M. T. Gastner and
the support from the Hungarian National Research, Development and Innovation Office NKFIH (K146736).

\section*{Data availability}
The fitted data currently available for a request from the authors.

\section{Supplementary Materials}

In this section, we present in detail the methods (i.e.binning, statistical fitting) employed in our analyses. 

\subsection{Data Binning}

\subsubsection{The Logarithmic and Linear Binning}
\begin{figure}[H]
    \centering
    \includegraphics[width=\textwidth]{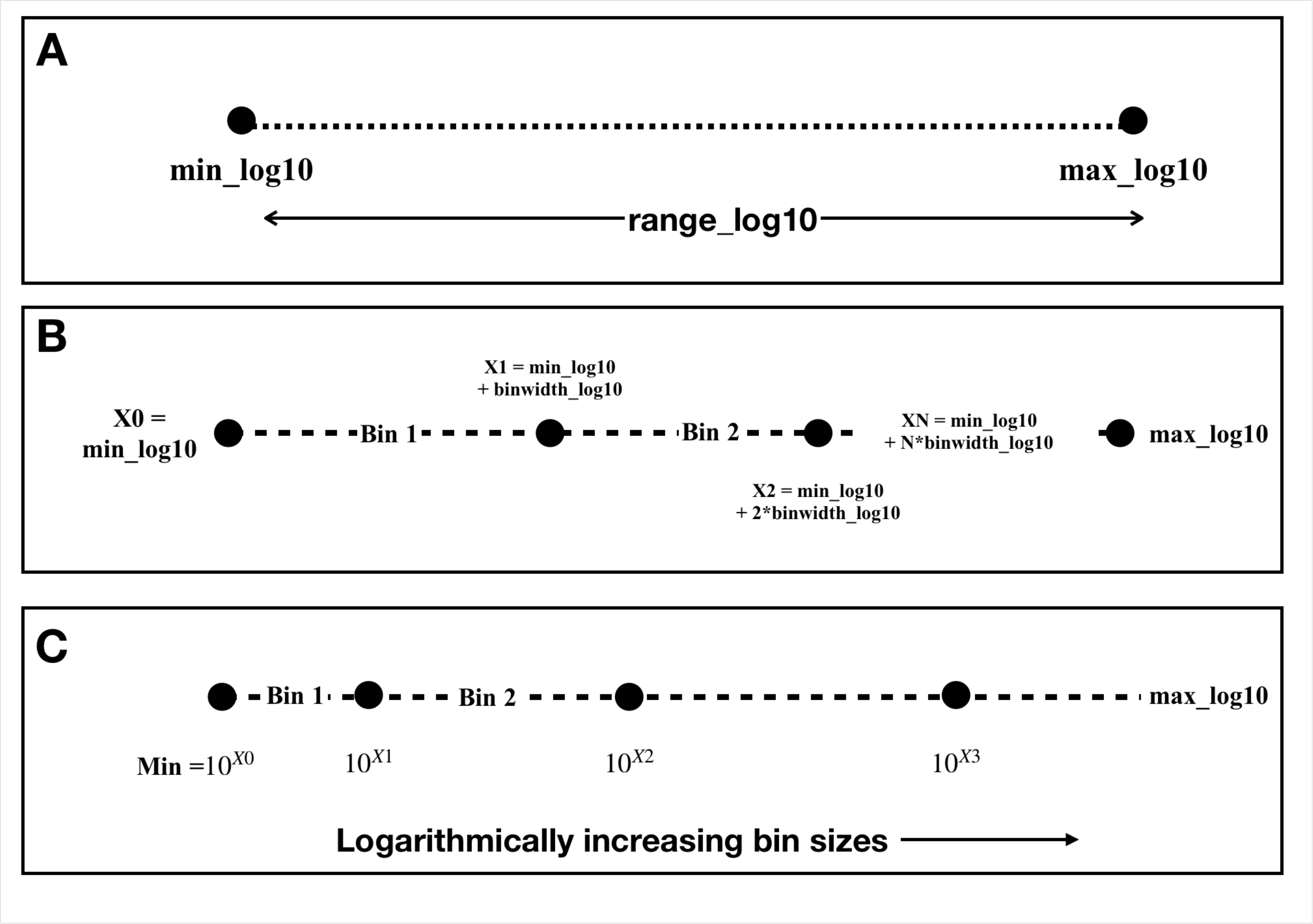}
    \caption{Schematic of Logarithmic Binning. (a) To be equally spaced in the log scale, the range of the data is defined by taking the difference of the log of the minimum and maximum of data (b) The range of data is divided by the number of bins set. The more data points, the larger the number of bins (c) How it looks like in the linear scale.}
    \label{fig:logbin}
\end{figure}

In this work, we employed both linear and logarithmic binning to our data. We did this because we are dealing with datasets of varying sizes. Figure~\ref{fig:logbin} shows a schematic of the log binning employed.

Since the datasets vary in size, the number of bins set also varies. In general, the number of bins set ensures that there are no empty bins to avoid erratic trends. The choice of binning, either linear or logarithmic, is based on the number of data points $N$ and maximum value. For the case of linear binning, the same schematics apply except that we don't take the log of the data points.

In the following, we listed the type of binning employed for every dataset presented in the main text.

\subsubsection{Global weights distribution}

\begin{table}[H]
\caption{Summary of global weights data and employed binning}
\centering
\begin{tabular}{|c|c|c|c|}
\hline
Dataset          & Number of datapoints, $N$ & Max. value & Type of binning \\ \hline
KKI-113          & 48,096,501                & 1,377.0     & Logarithmic     \\ \hline
KKI-18          & 46,524,003                &  854.0       & Logarithmic     \\ 
\hline
H01 Human (1mm$^3$) & 76, 004    &  89       & Logarithmic     \\ 
\hline
Mouse Retina     & 577,350                   & 29.0       & Linear          \\ \hline
Fly (Hemibrain)  & 3,413,160                 & 4,299.0     & Logarithmic     \\ \hline
Fly (Hemibrain reciprocated)  & 3,251,362    & 4,299     & Logarithmic     \\ 
\hline
Fly (Full brain) & 3,794,527                 & 2,358.0     & Logarithmic     \\ 
\hline
Fly (Full brain filtered) & 157,904 & 524.0  & Logarithmic     \\ 
\hline
Fly (Larva)      & 110,677                   & 121.0      & Linear          \\ \hline
\end{tabular}
\end{table}

\subsubsection{Local node strengths distribution}

\begin{table}[H]
\caption{Summary of node in-strength data and employed binning}
\centering
\begin{tabular}{|c|c|c|c|}
\hline
Dataset          & Number of datapoints, $N$ & Max. value & Type of binning \\ \hline
KKI-113          & 1,598,266                 & 4,977.0    & Logarithmic     \\ \hline
KKI-18           & 752,358                   & 38,243.0   & Logarithmic     \\ \hline
H01 Human (1mm$^3$) & 13,579    &  94       & Logarithmic     \\ 
\hline
Mouse Retina     & 1,076                     & 5,000.0    & Linear          \\ \hline
Fly (Hemibrain)  & 21,662                    & 2,708.0    & Logarithmic     \\ \hline
Fly (Hemibrain reciprocated)  & 16,804    & 4224.0     & Logarithmic     \\ 
\hline
Fly (Full brain) & 124,778                   & 23,036.0   & Logarithmic     \\ \hline
Fly (Full brain filtered) & 18,103 & 637.0  & Logarithmic     \\ 
\hline
Fly (Larva)      & 2,952                     & 210.0      & Linear          \\ \hline
\end{tabular}
\end{table}

\begin{table}[H]
\caption{Summary of node out-strength data and employed binning}
\centering
\begin{tabular}{|c|c|c|c|}
\hline
Dataset          & Number of datapoints, $N$ & Max. value & Type of binning \\ \hline
KKI-113          & 1,598,266                 & 5,010.0    & Logarithmic     \\ \hline
KKI-18           & 749,667                   & 55, 441.0  & Logarithmic     \\ \hline
H01 Human (1mm$^3$) & 13, 579    &  95.0       & Logarithmic     \\ 
\hline
Mouse Retina     & 1, 076                    & 6,880.0    & Linear          \\ \hline
Fly (Hemibrain)  & 21, 662                   & 5,044.0    & Logarithmic     \\ \hline
Fly (Hemibrain reciprocated)  & 16,804    & 4,378.0     & Logarithmic     \\ 
\hline
Fly (Full brain) & 124,778                   & 12,898.0   & Logarithmic     \\ \hline
Fly (Full brain filtered) & 18,103 & 557.0  & Logarithmic     \\ 
\hline
Fly (Larva)      & 2, 952                    & 160.0      & Linear          \\ \hline
\end{tabular}
\end{table}

\begin{table}[H]
\caption{Summary of node  total (in + out) node strength data and employed binning}
\centering
\begin{tabular}{|c|c|c|c|}
\hline
Dataset          & Number of datapoints, $N$ & Max. value & Type of binning \\ \hline
KKI-113          & 1,598,266                 & 6, 285.0   & Logarithmic     \\ \hline
KKI-18           & 797, 759                  & 73, 451.0  & Logarithmic     \\ \hline
H01 Human (1mm$^3$) & 13, 579    &  159.0       & Logarithmic     \\ 
\hline
Mouse Retina     & 1, 076                    & 7, 853.0   & Linear          \\ \hline
Fly (Hemibrain)  & 21, 662                   & 7,511.0    & Logarithmic     \\ \hline
Fly (Hemibrain reciprocated)  & 16,804    & 6,394.0  & Logarithmic     \\ 
\hline
Fly (Full brain) & 124,778                   & 26, 593.0  & Logarithmic     \\ \hline
Fly (Full brain filtered) & 18,103 & 1,194.0  & Logarithmic     \\ 
\hline
Fly (Larva)      & 2, 952                    & 331.0      & Linear          \\ \hline
\end{tabular}
\end{table}

\begin{table}[H]
\caption{Summary of source and sink node strength data and employed binning}
\begin{adjustbox}{width=\columnwidth,center}
\centering
\begin{tabular}{|l|ll|ll|l|}
\hline
\multicolumn{1}{|c|}{Dataset} &
  \multicolumn{2}{c|}{\begin{tabular}[c]{@{}c@{}}Sources\\ $s_{in}=0$\end{tabular}} &
  \multicolumn{2}{c|}{\begin{tabular}[c]{@{}c@{}}Sinks\\ $s_{out}=0$\end{tabular}} &
  \multicolumn{1}{c|}{Type of binning} \\ \hline
\multicolumn{1}{|c|}{} &
  \multicolumn{1}{c|}{No. of datapoints, $N$} &
  \multicolumn{1}{c|}{Max. value} &
  \multicolumn{1}{c|}{No. of datapoints, N} &
  \multicolumn{1}{c|}{Max. value} &
  \multicolumn{1}{c|}{} \\ \hline
KKI-113                      & \multicolumn{1}{l|}{6684} & 2110.0 & \multicolumn{1}{l|}{5,686} & 536   & Logarithmic \\ \hline
KKI-18                       & \multicolumn{1}{l|}{2270} & 1663.0 & \multicolumn{1}{l|}{2,410} & 1,300 & Logarithmic \\ \hline
H01 Human (1mm$^3$)          & \multicolumn{1}{l|}{2,109}  & 42     & \multicolumn{1}{l|}{2,145}  & 42    & Logarithmic \\ \hline
Mouse Retina                 & \multicolumn{1}{l|}{1}    & 3977.0 & \multicolumn{1}{l|}{61}    & 913   & Linear  (sinks only)    \\ \hline
Fly (Hemibrain)              & \multicolumn{1}{l|}{-}    & -      & \multicolumn{1}{l|}{46}    & 250   & Linear      \\ \hline
Fly (Hemibrain reciprocated) & \multicolumn{1}{l|}{3}     &     17.0   & \multicolumn{1}{l|}{1}      &    3   &      -       \\ \hline
Fly (Full Brain)             & \multicolumn{1}{l|}{8,565} & 640.0  & \multicolumn{1}{l|}{8,809} & 766   & Logarithmic \\ \hline
Fly (Full Brain filtered)             & \multicolumn{1}{l|}{2,023} & 31.0  & \multicolumn{1}{l|}{2,370} & 39.0   & Linear \\ \hline
Fly (Larva)                  & \multicolumn{1}{l|}{58}   & 18.0   & \multicolumn{1}{l|}{43}    & 38    & Linear      \\ \hline
\end{tabular}
\end{adjustbox}
\end{table}

\subsection{The Power-Law}

Functional brain networks have been proposed to exhibit scale-free behaviors with power-laws present~\cite{EguiluzPRL2005,HeuvelNEUROIM2008}. Power-laws indicate a specific degree of self-organization, either by growth and preferred attachment or replication, as in biological/metabolic networks~\cite{BarabasiSCIENCE1999,PiekniewskiIEEE2009}. When there is a transition from an ordered to an unordered phase without a characteristic length scale, power laws play a crucial role in statistical physics. Both theoretical and practical data support the notion that the brain functions near this important area~\cite{ShewNEUSCI2013, HaimoviciARXIV2013, HilgetagPHILOTRANS2000, BeggsJNEURO2003}.

\mage{An observable $x$ follows a power-law if it is drawn from the probability distribution}

\begin{equation}
\label{eqn:power-law}
P (x) \propto x^{-\alpha}
\end{equation} 

\noindent \mage{where $\alpha$ is a constant parameter of the distribution known as the exponent or scaling parameter. In reality, only few quantities follow a power-law for all values of $x$. Many times, the power-law applies only for values greater than some minimum $x_{min}$ (where $x_{min} \geq 0$) such that we say that the tail of the distribution follows a power law.} In this work, we fitted our data (Figure~\ref{fig:global} and Figure~\ref{fig:condprob}) by employing the methods of Clauset et.al.~\cite{ClausetSIAM2009} implemented by Alstott in Python~\cite{AlstottPLOS2014}.

\mage{Power-law distributions can either be continuous distributions governing continuous real numbers or discrete distributions where the quantity of interest can take only a discrete set of values, typically positive integers.}

\mage{As we are dealing with the number of edges emanating from a node, we utilize the power-law in the discrete case where we only consider integer values with probability distribution of the form,}

\begin{equation}
\label{eqn:discrete}
P (x) = Pr(X = x) = Cx^{-\alpha}
\end{equation} 

\mage{where $x_{min} > 0$.} More details of the normalization constant $C$ and specific form of the parameters $\alpha$ and $\sigma$ for the discrete power-law distribution case can be found in Ref.~\cite{ClausetSIAM2009}.

\mage{There are many ways to measure the distance between two probability distributions, but for non-normal data the most common is the Kolmogorov-Smirnov or KS statistic (Equation~\ref{eqn:KSD}), which is simply the maximum distance between the CDFs of the data and the fitted model~\cite{ClausetSIAM2009}. Note that, the KS statistic they used is different from KS test (which means comparing it with the Kolmogorov distribution). Here, the KS statistic served two purposes: (1)as a distance measure for fitting; wherein they estimate a value $\hat x_{min}$ that minimizes $D$ (2) distance measure to test goodness of fit, by using bootstrapping.  Instead of using the KS distribution as the distribution for the KS statistic under the null hypothesis, one estimates an empirical distribution for this statistic by simulations. Here, they generated  a large number of power-law distributed synthetic data sets with scaling parameter $\alpha$ and lower $x_{min}$ equal to those of the distribution that best fits the observed data. This distance is compared with distance measurements for comparable synthetic data sets drawn from the same model, and the p-value is defined to be the fraction of the synthetic distances that are larger than the empirical distance.
}

\begin{equation}
\label{eqn:KSD}
D_{\alpha} = \max_{x \geq x_{min}} \, |S(x) - P(x)|
\end{equation} 

\noindent \mage{where $S(x)$ is the CDF of the data for the observations with value at least $x_{min}$ and $P(x)$ is the CDF for the power-law model that best fits the data in the region $x \geq x_{min}$. The estimated value of $\hat x_{min}$ is then the value of $x_{min}$ that minimizes $D$}. Finally, we also computed the standard error $\sigma$ of our estimate for the power-law exponent.

\subsection{The Stretched Exponential}
To account quantitatively for many reported natural fat tail distributions in nature and economy, Laherrere and Sornette~\cite{LaherreEPJB1998} proposed the stretched exponential~\ref{eqn:stretched} as an alternative to the power-law. 

\begin{equation}
\label{eqn:stretched}
P (x) dx = c \left( \frac{x^{c-1}}{A_o^{c}}\right) e^{\left[(\frac{-x}{A_o})^c \right]}  dx
\end{equation} 

such that the cumulative distribution is 

\begin{equation}
\label{eqn:stretched_cdf}
P_{c}(x) = e^{\left[(\frac{-x}{A_o})^c \right]}
\end{equation} 

Stretched exponentials are characterized by an exponent $c < 1$, in which the exponent $c$ is the inverse of the number of generations (or products) in a multiplicative process. The borderline $c=1$ corresponds to the usual exponential distribution. For $c < 1$, the distribution~\ref{eqn:stretched_cdf} presents a clear curvature in a log-log plot while exhibiting a relatively large apparent linear behavior, all the more so, the smaller $c$ is. It can thus be used to account both for a limited scaling regime and a cross-over to non-scaling. When using the stretched exponential pdf, the rationale is that the deviations from a power law description are fundamental and not only a finite-size correction.

Among its numerous benefits is its economy—it has only two movable parameters with definite physical meaning. Moreover, it originates from a straightforward and universal mechanism concerning multiplicative processes.

To find the fitting for the in, out, and total strength distributions, here I computed for the best fitting parameters $A_o$ and $c$ (exponent) of the stretched exponential by using the mean and standard deviation of the data and scanning through a range of values for $c$ (from 0.1 to 0.9999).

Adapting from the Appendix section of~\cite{LaherreEPJB1998}, this section shows the derivation of the condition for the data to follow a stretched exponential. We start with the mean of the stretched exponential~\ref{eqn:stretched} given by 

\begin{equation}
\label{eqn:stretched_mean}
<x> = x_o \left( \frac{1}{c_o}\right) \Gamma \left( \frac{1}{c_o}\right)
\end{equation} 

and its variance is

\begin{equation}
\label{eqn:stretched_var}
\sigma^2 = x_o^{2} \left( \frac{2}{c_o}\right) \Gamma \left( \frac{2}{c_o}\right) - <x>^{2}
\end{equation} 

From ~\ref{eqn:stretched_mean}:

\begin{equation}
\label{eqn:stretched_xo}
x_o = \frac{c_o<x>}{\Gamma \left( \frac{1}{c_o}\right)}
\end{equation} 

Plug-in~\ref{eqn:stretched_xo} to ~\ref{eqn:stretched_var} we get:

\begin{equation}
\label{eqn:stretched_var_exp}
\sigma^2 = \frac{c_o^2<x>^2}{\Gamma^2 \left( \frac{1}{c_o}\right)} \left( \frac{2}{c_o}\right) \Gamma \left( \frac{2}{c_o}\right) - <x>^{2}
\end{equation} 

Simplifying~\ref{eqn:stretched_var_exp}  by factoring out $<x>$ we get, 

\begin{equation}
\label{eqn:stretched_var_exp2}
\sigma^2 = \left[ 2c_o \frac{\Gamma \left( \frac{2}{c_o}\right)}{\Gamma \left( \frac{1}{c_o}\right)} - 1 \right] <x>^2
\end{equation} 

Transpose $<x>^2$ to the other side, 

\begin{equation}
\label{eqn:stretched_var_exp3}
\frac{\sigma^2}{<x>^2} = \left[ 2c_o \frac{\Gamma \left( \frac{2}{c_o}\right)}{\Gamma \left( \frac{1}{c_o}\right)} - 1 \right] 
\end{equation} 

Add 1 to both sides of~\ref{eqn:stretched_var_exp3},

\begin{equation}
\label{eqn:stretched_var_exp4}
\frac{\sigma^2}{<x>^2} + 1 = \left[ 2c_o \frac{\Gamma \left( \frac{2}{c_o}\right)}{\Gamma \left( \frac{1}{c_o}\right)} \right] 
\end{equation} 

Divide everything by 2 and rearrange,

\begin{equation}
\label{eqn:stretched_var_exp5}
\left[ c_o \frac{\Gamma \left( \frac{2}{c_o}\right)}{\Gamma \left( \frac{1}{c_o}\right)} \right]  = \frac{\sigma^2}{2<x>^2} + \frac{1}{2}
\end{equation} 

If we let the LHS of~\ref{eqn:stretched_var_exp5} be, 

\begin{equation}
\label{eqn:stretched_var_exp6}
F_o = \left[ c_o \frac{\Gamma \left( \frac{2}{c_o}\right)}{\Gamma \left( \frac{1}{c_o}\right)} \right] 
\end{equation}

and the RHS be some constant $F_o$ which is a function of $\sigma$,  the standard deviation and $<x>$ is the mean of the data.

\begin{equation}
\label{eqn:stretched_var_exp7}
F = \frac{\sigma^2}{2<x>^2} + \frac{1}{2}
\end{equation}

Here, if the conditions $f = |F - F_o| = 0$ and $c < 1$ are met, then the data follow a stretched exponential trend. 

\subsection{The Exponential Truncated Power-Law}

In this work, we used the exponential truncated power-law of the form

\begin{equation}
\label{eqn:Exp_TPL}
P(x) = C (x + x_o)^{-\beta} \; \exp(-x/ \kappa)
\end{equation}

\noindent from the work of Gonzalez and Barab\'asi~\cite{GonzalezNATURE2008}. Simply put, an exponentially truncated power-law is a power law multiplied by an exponential function. Since they applied this to displacements $r$ in cities, we replaced the variable $r$ with $x$ to make it into a more general form. Here, the parameter $\beta$ is the power-law exponent that is valid for small values of $s$, and $\kappa$ is the cut-off value. To determine the fitting parameters $s_o$, $\beta$, and $\kappa$, we \mage{employed the {\tt curve\_fit()} function from Python's scipy.stats \cite{VirtanenNATUREMETH2020}. We used both Scipy version 1.10.1 with LAPACK 0.3.18 and Scipy version 1.13.1 with LAPACK 0.3.27}. To check the goodness-of-fit of our data to the said distribution, we compute for the $R^2$ value, defined as

\begin{equation}
\label{eqn:R2}
R^2 = 1 - \frac{SS_{res}}{SS_{tot}}
\end{equation}

where $SS_{res}$ is the residual sum of squares

\begin{equation}
\label{eqn:SSres}
SS_{res} = \sum_i (y_i - f_i)^2
\end{equation}

where $y_i$ is the observed data associated with a fitted or predicted value $f_i$. On the other hand, the total sum of squares $SS_{tot}$ is given by

\begin{equation}
\label{eqn:SStot}
SS_{res} = \sum_i (y_i - \bar y)^2
\end{equation}

\noindent where $\bar y$ is the mean of the data.

\subsection{The Lognormal Distribution}

Based on earlier research~\cite{SongPLOS2005}, synaptic strengths in the rat visual cortex follow a lognormal distribution with a heavy tail, indicating a higher-than-expected abundance of strong synaptic connections. Here, we opted to fit the node strength distributions of the mouse retina with a lognormal distribution as well.

The lognormal distribution is a continuous probability distribution of a random variable whose logarithm is normally distributed. A lognormal process is the statistical realization of the multiplicative product of many independent random variables, each of which is positive. 

\begin{equation}
\label{eqn:lognormal}
P(x) = \frac{1}{x\sigma \sqrt{2\pi}} \exp \left(\frac{-(ln \; x - \mu)^2}{2\sigma^2}\right)
\end{equation}

\noindent where $\mu$ is the expected value (or mean) and $\sigma$ is the standard deviation of the variable's natural logarithm, not the expectation and standard deviation of observed variable $x$ itself.

\begin{equation}
\label{eqn:mu}
\mu = \ln \; \left( \frac{\mu_x^2}{\sqrt{\mu_x^2 + \sigma^2}}       \right)
\end{equation}

\begin{equation}
\label{eqn:sigma}
\sigma = \ln \; \left( 1 + \frac{\sigma_x^2}{\mu_x^2} \right)
\end{equation}

\noindent where $\mu_x$ and $\sigma_x$ are the mean and standard deviation of the data.

\bibliographystyle{unsrt}
\bibliography{references.bib}

\end{document}